\documentclass[10pt,journal,twocolumn,twoside]{IEEEtran-v17}
\usepackage{epsfig,amsfonts,amsbsy,bm,mathrsfs}
\usepackage[nolist]{acronym}
\usepackage{mathrsfs} 
\usepackage{graphicx,cite,amssymb,amsmath,bbold}
\usepackage{mathtools}
\mathtoolsset{showonlyrefs}

\usepackage{tikz}
\usetikzlibrary{fadings}
\usetikzlibrary{shadows.blur}
\usetikzlibrary{shapes,arrows}
\usetikzlibrary{calc,shapes.misc}
\usetikzlibrary{decorations.pathreplacing}
\usepackage{ifthen}
\usetikzlibrary{positioning}
\usepackage{pgfplots,relsize}

 \newcommand{\lefto}{\mathopen{}\left}
\begin{document} 
\newcommand{\vecy}{\bm{y}}
\newcommand{\vecY}{\bm{Y}}
\newcommand{\vecx}{\bm{x}}
\newcommand{\vecX}{\bm{X}}
\newcommand{\vecn}{\bm{n}}
\newcommand{\vecN}{\bm{N}}
\newcommand{\preamb}{\bm{s}}
\newcommand{\info}{\bm{u}}
\newcommand{\parity}{\bm{p}}
\newcommand{\csi}{h}
\newcommand{\CSI}{H}
\newcommand{\estcsi}{\hat{h}}
\newcommand{\estCSI}{\hat{H}}
\newcommand{\err}{\delta}
\newcommand{\ERR}{\Delta}
\newcommand{\CN}{\mathcal{CN}}
\newcommand{\var}{\sigma^2}
\newcommand{\errvar}{\sigma_\ERR^2}
\newcommand{\perr}{\theta}
\newcommand{\Perr}{\Theta}
\newcommand{\argmax}[1]{\underset{#1}{\mathrm{arg \, max}}\,}
\newcommand{\code}{\mathscr{C}}
\newcommand{\mm}{{q}}
\newcommand{\capacity}{{\mathrm{C}}}
\newcommand{\mi}{{\mathrm{I}}}
\newcommand{\gmi}{{\mathrm{I}}^\mm}
\newcommand{\inalpha}{\mathcal{X}}
\newcommand{\outalpha}{\mathcal{Y}}
\newcommand{\de}{\mathrm{d}}
\newcommand{\avg}{\mathsf{E}}

\newtheorem{mydef}{Definition}
\newtheorem{prop}{Proposition}
\newtheorem{theorem}{Theorem}
\newtheorem{lemma}{Lemma}
\newtheorem{example}{Example}
\newtheorem{corollary}{Corollary}

\renewcommand{\Re}{\mathbb{Re}}
\renewcommand{\Im}{\mathbb{Im}}

\begin{acronym}
\acro{AWGN}{additive white Gaussian noise}
\acro{bi-AWGN}{binary input additive white Gaussian noise}
\acro{BPSK}{binary phase shift keying}
\acro{CSI}{channel state information}
\acro{i.i.d.}{independent and identically distributed}
\acro{IRA}{irregular repeat accumulate}
\acro{LDPC}{low-density parity-check}
\acro{MAP}{maximum a posteriori}
\acro{ML}{maximum likelihood}
\acro{PEG}{progressive edge growth}
\acro{QAM}{quadrature amplitude modulation}
\acro{r.v.}{random variable}
\acro{SPC}{single parity check}
\acro{GRCB}{Gallager's random coding bound}
\acro{SPB}{sphere packing bound}
\acro{LLR}{log-likelihood ratio}
\acro{RHS}{right-hand side}
\end{acronym} 

\title{Short Codes with Mismatched Channel State Information: A Case Study} 
 \author{\IEEEauthorblockN{Gianluigi Liva$^1$, Giuseppe Durisi$^2$, Marco Chiani$^3$, Shakeel Salamat Ullah$^4$, Soung Chang Liew$^4$\\} \IEEEauthorblockA{ $^1$Deutsches Zentrum f\"ur Luft- und Raumfahrt (DLR), We{\ss}ling, Germany\\
 $^2$Chalmers University of Technology, Gothenburg, Sweden\\
 $^3$University of Bologna, Bologna, Italy\\ $^4$Chinese University of Hong Kong, Hong Kong, China}}

\maketitle
\begin{abstract}
  The rising interest in applications requiring the transmission of small amounts of data has recently lead to the development of accurate performance bounds and of powerful channel codes for the transmission of short-data packets over the AWGN channel.
  Much less is known about the interaction between error control coding and channel estimation at short blocks when transmitting over channels with states (e.g., fading channels, phase-noise channels, etc\dots) for the setup where no \emph{a priori} channel state information (CSI) is available at the transmitter and the receiver.
  In this paper, we use the mismatched-decoding framework to characterize the fundamental tradeoff occurring in the transmission of  short data packet over an AWGN channel with unknown gain that stays constant over the packet.
  Our analysis for this simplified setup aims at showing the potential of mismatched decoding as a tool to design and analyze transmission strategies for short blocks. We focus on a pragmatic approach where the transmission frame contains a codeword as well as a preamble that is used to estimate the channel (the codeword symbols are not used for channel estimation).
  Achievability and converse bounds on the block error probability achievable by this approach are provided and compared with simulation results for schemes employing short low-density parity-check codes. Our bounds turn out to predict accurately the optimal trade-off between the preamble length and the redundancy introduced by the channel code.
\end{abstract}

\section{Introduction}\label{sec:intro}   
The need for machine-type communications and for telecommand and remote control systems that operate under strict latency and reliability constraints has recently caused a rising interest in protocols and error correction schemes for the transmission of small amounts of data~\cite{Durisi16:Short}.
Considerable efforts have been spent over the last years in the design of powerful short error-correcting codes (see, e.g.,~\cite{Liva2016:ShortSurvey} and the references therein), and in understanding the fundamental performance limits in the regime of small block size \cite{Dolinar98:BOUNDS,Fossorier04:BOUNDS,SasonShamai06:BOUNDS,Polyanskiy10:BOUNDS}.
Much less is known about the interaction between error control coding and other crucial receiver operations such as synchronization and channel estimation. 
For instance, very few works study the rates achievable with short codes over fading channels, for a given packet error probability requirement, when \ac{CSI} is not available \emph{a priori} at the receiver and at the transmitter \cite{yang14-07c,durisi2016short,Ostman2017:URLLC}.
Furthermore, most of the available achievability bounds rely upon \emph{noncoherent} transmission strategies and do not target explicitly the practically relevant setup in which pilot symbols are embedded within each transmission frame to enable channel estimation at the decoder.

Information theoretic tools that are useful in investigating this specific setup are those relying on \emph{mismatched decoding}---a framework that allows one to characterize performance limits when the decoding rule is not matched to the statistical law governing the communication channel \cite{MM:Kaplan93,MM:Merhav94,MM:Csiszar95,MM:Lapidoth96,MM:Ganti00,MM;Asyhari14,scarlett14-a}. 
In particular, the mismatch may be caused by an imperfect estimation of the channel state based on pilot transmission, which may lead to the adoption of a suboptimal decoding metric.


A thorough understanding of this problem and of the involved fundamental tradeoffs is of paramount importance to design efficient communication protocols for the transmission of short messages. 
Indeed, when packets are large, a considerable amount of channel uses can be dedicated to pilot transmission, which allows the decoder to acquire almost perfect \ac{CSI} (provided that the channel state varies sufficiently slowly), without affecting in a tangible manner the transmission rate. 
On the contrary, when packets are short the use of a large number of pilots may yield an unacceptable rate loss. 
This calls for a precise analysis of the tradeoff between the number of channel uses dedicated to estimating the channel, and the number of channel uses allocated to the transmission of the coded information at short block lengths. 

The aim of this paper is to demonstrate, through a simple yet practically relevant example, the usefulness of  mismatch decoding  as a tool to design and analyze transmission strategies for short blocks.
Specifically, we shall address the problem of transmitting a short data packet over an \ac{AWGN} channel with unknown (complex) channel coefficient that stays constant over the duration of the transmission frame.

%
%
We focus on the pragmatic approach  where the transmission frame is split into two fields: a field hosting pilot symbols (\emph{preamble}) and a field containing the encoded information (\emph{data field}). 
Furthermore, we focus on the case where only the preamble is used to estimate the channel (i.e., no use is made of the data field for channel estimation purposes). 
This pragmatic approach is likely far from optimal, as recently exemplified in \cite{Ostman2017:URLLC}, but it is prevalent in practical implementations.
We provide achievability and converse bounds on the block error probability achievable by this approach. 
Furthermore, we compare our bounds with simulation results for schemes employing short \ac{LDPC} codes \cite{Gallager63:LDPC}, and show that the bounds allow one to accurately predict the optimal trade-off between the size of the preamble and the redundancy introduced by the channel code.

\section{Preliminaries}\label{sec:prelim}
In the following, random variables and their realizations are denoted by uppercase and lowercase letters, respectively. 
We consider the transmission of $k$ bits of information over $N$ channel uses of the complex \ac{AWGN} channel
\begin{equation}
  Y_\ell=\csi X_\ell+Z_\ell, \quad \ell=1,\ldots,N.  \label{eq:channel}
\end{equation}
Here, the input symbols $\{X_\ell\}$ are assumed to belong to a finite cardinality constellation $\inalpha \subset \mathbb{C}$, whose average power is normalized to one, i.e., $|\inalpha|^{-1}\sum_{x \in \inalpha}|x|^2=1$. The noise samples are independent and $\CN(0,2\var)$-distributed and the channel coefficient $h$ is complex, it is constant over the frame (i.e., over the $N$ channel uses), and it is not known at the transmitter and at the receiver.

\subsection{Pragmatic Approach}
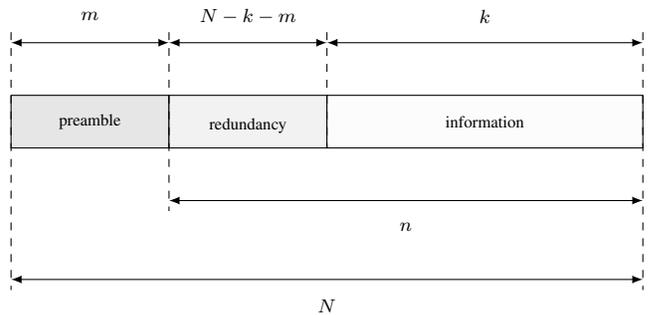
\begin{figure}
	\begin{center}
			\begin{tikzpicture}[
	myrect/.style={
		rectangle,
		draw,
		thin,
		inner sep=2pt,
		fit=#1},
	scale=0.7, 
	every node/.style={scale=0.9}
	]
	
	\coordinate (A) at (-4,0);
	\coordinate (B) at (-1,1);
	\draw[fill=gray!20] (A) rectangle (B) node[pos=.5] {\scriptsize preamble};

	\coordinate (C) at (2,0);
	\draw[fill=gray!10] (B) rectangle (C) node[pos=.5,yshift=-0.06cm] {\scriptsize redundancy};
	
	\coordinate (D) at (8,1);
	\draw[fill=gray!2] (C) rectangle (D) node[pos=.5] {\scriptsize information};

	\draw[dashed] (-4,2.2) -- (-4,-2.7);
	\draw[dashed] (-1,2.2) -- (-1,-1.2);
	\draw[dashed] (2,2.2) -- (2,0);
	\draw[dashed] (8,2.2) -- (8,-2.7);	
	
	\draw[latex-latex] (-4,2) -- (-1,2);
	\draw[latex-latex] (-1,2) -- (2,2);
	\draw[latex-latex] (2,2) -- (8,2);
	\draw[latex-latex] (-1,-1) -- (8,-1);
	\draw[latex-latex] (-4,-2.5) -- (8,-2.5);
	
	\node at (-2.5,2.5) {\footnotesize $m$};
	\node at (0.5,2.5) {\footnotesize$N-k-m$};
	\node at (5,2.5) {\footnotesize$k$};
	\node at (3.5,-1.5) {\footnotesize$n$};
	\node at (2,-3) {\footnotesize$N$};

	\end{tikzpicture}
	\end{center}
	\caption{Frame structure (pragmatic approach) for a binary systematic channel code: the $N$ channel uses are divided among \emph{preamble} ($m$ symbols), \emph{redundancy} ($N-k-m$ symbols), and \emph{information} ($k$ symbols).}
	\label{fig:structure}
\end{figure}

We consider a pragmatic approach where $m$ out of the $N$ channel uses are employed to transmit a preamble containing pilot symbols known to the receiver.
The remaining $n=N-m$ channel uses are employed to  transmit the $k$ data symbol, which are encoded using an $(n,k)$ code. 
We refer to the overall rate of this scheme  as $R=k/N$, whereas the code rate is denoted by $R_c=k/n$.  
An example of the frame structure just described, for the case of a binary coding scheme with systematic encoding, is depicted in Fig.~\ref{fig:structure}.

The receiver estimates the channel coefficient using the preamble, and the channel estimate is provided to the decoder of the $(n,k)$ code, which treats the estimate as if it was perfectly accurate (mismatch decoding). 
When designing such a scheme, one faces the following trade-off between channel estimation and channel coding: for  given $k$ and $N$, one may decide to acquire an accurate channel estimate by taking $m$ large, which, however, implies also choosing a weak channel code, i.e, one that introduces a low redundancy and has a large code rate. 
Alternatively, one may decide to accept a less accurate channel estimate by taking $m$ small and to utilize a more robust channel code, i.e., one that introduces more redundancy and has a smaller code rate. 
The purpose of this paper is to shed lights on this tradeoff.

\subsection{Performance with Ideal CSI}
Without loss of generality, we shall assume throughout the reminder of the paper that the unknown deterministic channel coefficient has value $h=1$.  In this case, the signal-to-noise ratio is $(2\sigma)^{-1}$ and  $E_b/N_0=(2R\var)^{-1}$ where $E_b$ denotes the energy per information bit, and $N_0$ is the single-sided noise power spectral density. 
The channel transition probability density function is the complex Gaussian
\[
W(y|x;h)=\frac{1}{2\pi\var}\exp \lefto(-\frac{1}{2\var}\left|y-hx\right|^2\right)
\] 
and (in the ideal \ac{CSI} setting) it is perfectly known to the receiver.
In the asymptotic limit of large frame size (i.e., large $N$), reliable transmission can be achieved by selecting a code with rate $R_c$ lower than the channel capacity $\log(1+|h|^2/(2\var))$ independently of whether $h$ is known to the receiver or not. Indeed, since $h$ is assumed to be constant over the frame (independently of its length), one can perfectly estimate the channel at the receiver with a negligible rate penalty.
As we shall see later, this is no longer the case when the frame is short.

As performance metric for the case of short frames, we shall use \ac{GRCB} \cite{Gallager68:BOOK}, which---for the case of perfect CSI---gives the following upper bound to the minimum average block error probability $P^*_B$ achievable using  $(n,k)$ codes:
\[
P^*_B\leq \bar{P}_B
\]
with
\[
\bar{P}_B = 2^{-n E_G(R_c)} 
\]
and
$\displaystyle
E_G(R_c)=\max_{0\leq \rho \leq 1} \left(E_0(\rho)-\rho R_c\right)
$
where
\begin{equation}\label{eq:gallager_function_perfect_CSI}
E_0(\rho)=-\log_2 \avg \lefto[\left(\avg \lefto[ \left(\frac{W(Y|X';h)}{W(Y|X;h)}\right)^{\frac{1}{1+\rho}}  \Bigg|X,Y \right] \right)^\rho\right].
\end{equation}
Here, $(X,Y,X')\sim Q(x)W(y|x;h)Q(x')$, and we choose the input distribution $Q(\cdot)$ to be the uniform distribution over the constellation alphabet $\mathcal{X}$.
Writing the Gallager's $E_0$  function in the specific form given in~\eqref{eq:gallager_function_perfect_CSI} will turn out useful when we introduce the mismatched case.

%
%
%
%
  
\section{Pragmatic Approach: Performance Benchmarks under Mismatched Decoding}\label{sec:benchmarks_pragmatic}

In this section, by leveraging on the mismatched decoding approach \cite{MM:Kaplan93,MM:Csiszar95}, we show how the \ac{GRCB} can be extended to provide an upper bound on the average block error probability $P_B^*$ for the case when the receiver acquires an  imperfect estimate of the channel coefficient $h$ through the pilot symbols contained in the preamble. 
We will then particularize the obtained result for the special case of \ac{BPSK} modulation, for which we will provide also a converse result, i.e., a lower bound on the block error probability.

We shall denote the \ac{ML} estimate of the unknown deterministic channel coefficient $h$ by the \ac{r.v.} $\estCSI$. 
Furthermore, we let $\estcsi$ be a realization of~$\estCSI$. 

%
The estimation error $\ERR=\csi-\estCSI$ is a  $\CN(0,\errvar)$-distributed \ac{r.v.} with $\errvar=2\var /m$. 
Finally, we denote the phase estimation error by $\perr=\arg(\estcsi)$.

\subsection{Random Coding Bound under Mismatched Decoding}
In the mismatched decoding framework, the receiver decodes the channel outputs $\{y_\ell\}_{\ell=1}^N$ using the following maximum metric rule (here, $\code$ is the set of codewords of the chosen code)
\[
\hat{\vecx}=\argmax{\vecx \in \code} \prod_{\ell=1}^n \mm(x_\ell,y_\ell).
\]
where $\mm(x,y)$ denotes the \emph{mismatched metric} and $\vecx=[x_1,\dots,x_N]$.
In our setting, the mismatched metric that results by treating the channel estimate $\estcsi$ as perfect, is $\mm(x,y,\estcsi)=W(y|x;\estcsi)$, i.e.,
\[
\mm(x,y,\estcsi)=\frac{1}{2\pi\var}\exp \Bigl(-\frac{1}{2\var}\bigl|y-\estcsi x\bigr|^2\Bigr).
\]
It follows from~\cite{MM:Kaplan93} that for a given channel estimate $\estcsi$, we have
\begin{equation}
P^*_B(\estcsi)\leq \bar{P}_B(\estcsi) \label{eq:RCB_forH}
\end{equation}
where
\begin{equation}
\bar{P}_B(\estcsi):= 2^{-n E_G(R_c,\estcsi)} \label{eq:RCB_forH_explic}
\end{equation}
with 
$\displaystyle
E_G(R_c,\estcsi)=\max_{0\leq \rho \leq 1} \sup_{s\geq 0}\, \lefto(E_0(s,\rho,\estcsi)-\rho R_c\right)
$
and
\begin{equation}
  \label{eq:mismatch_error_exponent}
E_0(s,\rho,\estcsi)=
 -\log_2 \avg \lefto[\left(\avg \lefto[ \left(\frac{\mm(X',Y,\estcsi)}{\mm(X,Y,\estcsi)}\right)^s  \Bigg|X,Y \right] \right)^\rho\right].
\end{equation}
Averaging over the random channel estimate, we finally get
\begin{equation}
\avg\lefto[P_B^*(\hat{H})\right]\leq \avg [\bar{P}_B(\hat{H})] =: \bar{P}_B\label{eq:RCB_AVG}
\end{equation}
Observe that \eqref{eq:RCB_AVG} guarantees the existence of a $(n,k)$ code with an average error probability lower than  the \ac{RHS} of \eqref{eq:RCB_AVG}, where the average is both over the noise and over the channel estimate, but not the existence of a $(n,k)$ code with error probability lower than the \ac{RHS} of \eqref{eq:RCB_forH} for \emph{every} $\hat{h}$.

\subsection{The BPSK Case}
Assume \ac{BPSK} modulation, i.e., that the channel input alphabet is $\inalpha=\left\{-1,+1\right\}$.
Denote the mismatch \emph{log-metric ratio} by
\[
L(Y,\estcsi)\triangleq\log\lefto[ \frac{\mm(+1,Y,\estcsi)}{\mm(-1,Y,\estcsi)}\right].
\]
Then
\begin{equation}
\avg \lefto[ \left(\frac{\mm(X',Y,\estcsi)}{\mm(X,Y,\estcsi)}\right)^s  \Bigg|X,Y \right]=
\frac{1}{2}+\frac{1}{2} \exp\lefto(-sXL(Y,\estcsi)\right). \label{eq:LLR_in_exp}
\end{equation}
We can write the log-metric ratio as
\begin{align}
L(Y,\estcsi)&=\frac{2}{\var}\Re\left\{Y\estcsi^*\right\}\\
&=\frac{2}{\var}|\estcsi|\cos(\perr) X + \frac{2}{\var}\Re\left\{Z|\estcsi|\exp(-j\perr)\right\}. 
\end{align}
Let $
Z'\triangleq \Re\{Z|\estcsi|\exp(-j\perr)\}$.
We next observe that $Z'$ is normally distributed with zero mean and variance $|\estcsi|^2 \var$.
It follows then from~\eqref{eq:LLR_in_exp} that the mismatch causes a scaling of the log-metric ratio by $|\estcsi|/\cos(\perr)$.
 This implies that, when $\cos(\perr)\geq 0$, the mismatched \ac{GRCB} is a function only of the phase mismatch $\perr$ between $h$ and $\estcsi$ and it does not depend on the amplitude mismatch.
 Indeed, let $\tilde{X}=X\cos(\perr)$, set $s=s' \cdot \cos(\perr)/|\estcsi|$ in~\eqref{eq:LLR_in_exp}, and then substitute~\eqref{eq:LLR_in_exp} in~\eqref{eq:mismatch_error_exponent}. 
 It follows from Holder's inequality \cite{Gallager68:BOOK} that the optimal value of $s'$ is  $s'=1/(1+\rho)$.
This implies that, for a given phase estimation error $\perr \in [-\pi/2,\pi/2]$  the mismatched \ac{GRCB} reduces to the random coding bound~\eqref{eq:gallager_function_perfect_CSI} of a binary-input real-valued AWGN channel with input $\tilde{X}$ and degraded signal-to-noise ratio 
\[(E_b/N_0)'=(E_b/N_0)+10\log_{10} \cos^2(\perr).\]
For $\perr\not\in[-\pi/2,\pi/2]$, the error probability is set to $1$ (see Example 5 in \cite{MM:Merhav94}).



In fact, one can apply steps similar to the ones just outlined to transform any available finite-blocklength bound  for the real AWGN channel that holds under the assumptions of \ac{ML} decoding and equal-power channel input vectors (this last assumption holds for any code combined with \ac{BPSK} modulation), into a finite-blocklength bound for our pragmatic scheme.
In particular, we can obtain a lower bound on the average error probability $P^{*}_B$ by using the 1959 \ac{SPB} \cite{Shannon59:SPB} for each phase estimation error $\perr$, and then by averaging over $\theta$.

\section{Numerical Results}

\begin{figure}
	\begin{center}
	 \includegraphics[width=1\columnwidth]{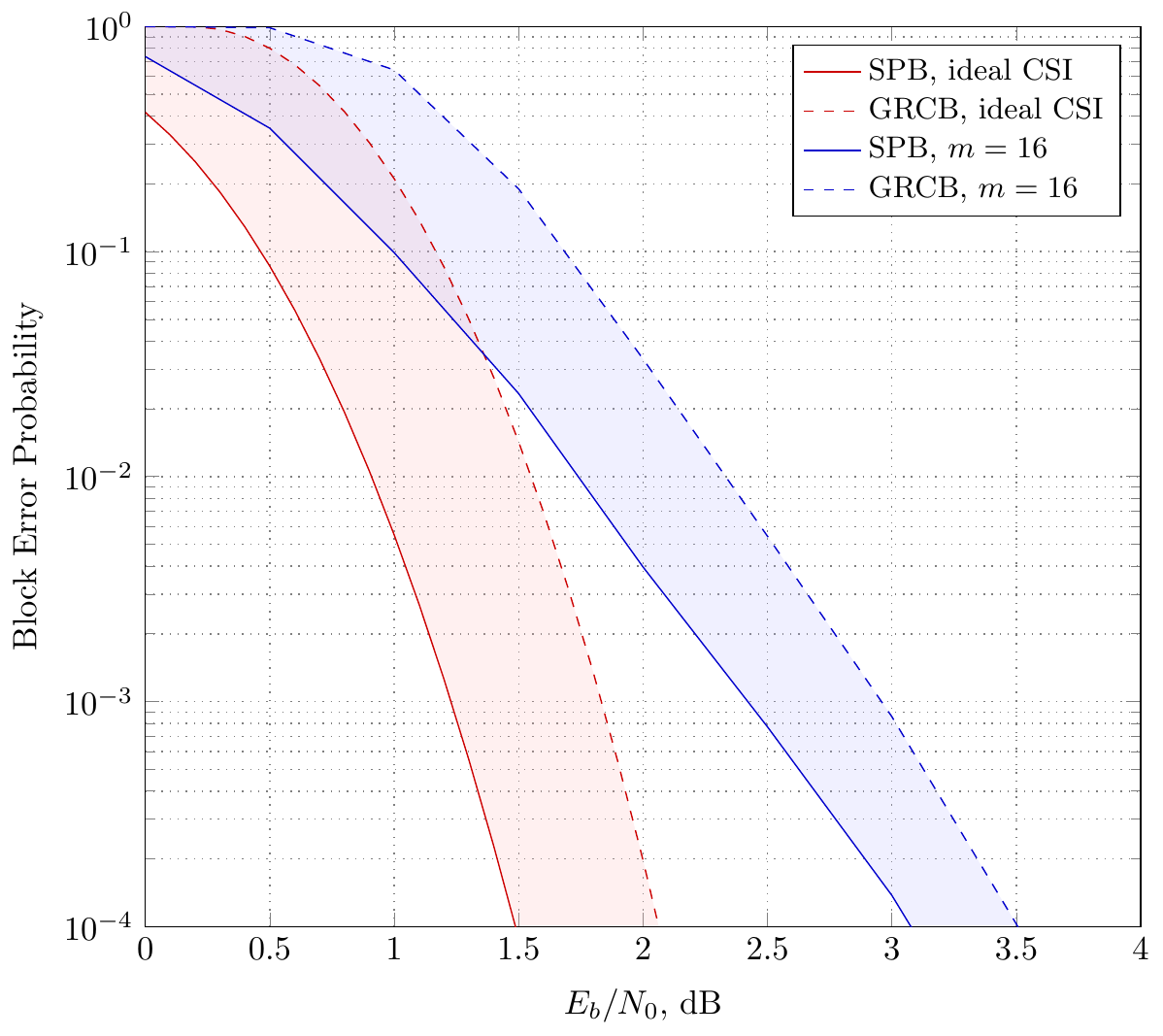}
	\end{center}
\caption{Block error probability vs. signal-to-noise ratio according to the \ac{SPB} and the \ac{GRCB} with ideal CSI and with pilot-based channel estimation. \ac{BPSK}  modulation, frame length $N=512$ symbols, $k=256$ information bits.}	\label{fig:PER_BPSK}
\end{figure}

In Fig.~\ref{fig:PER_BPSK}, we plot the \ac{SPB} and the \ac{GRCB} for the case of perfect~\ac{CSI} and BPSK modulation.
Here,~$N=512$ and $k=256$, which results in a overall rate $R=1/2$. 
Since we assumed perfect \ac{CSI}, there is no preamble ($n=N$) and the overall rate $R$ coincides with the code rate $R_c$. 
We also provide bounds for the case when the channel coefficient is estimated through a preamble of length $m=16$. 
Here, the code rate is increased to $R_c=256/(512-16)\approx 0.516$ to accommodate the $16$-symbol preamble. 
The gap between the \ac{GRCB} for the ideal \ac{CSI} case and the one for our pragmatic scheme increases as the block-error probability is decreased and it reaches about $1.5$ dB at a block error probability of $10^{-4}$. 
The two \ac{SPB} curves behave similarly.
Our bounds allow us to conclude that the block error probability curve of the best $(496,256)$ code for our pragmatic scheme (with preamble of size $m=16$) lies within the blue shaded area in Fig.~\ref{fig:PER_BPSK}. 

\begin{figure}
	\begin{center}
	\includegraphics[width=1\columnwidth]{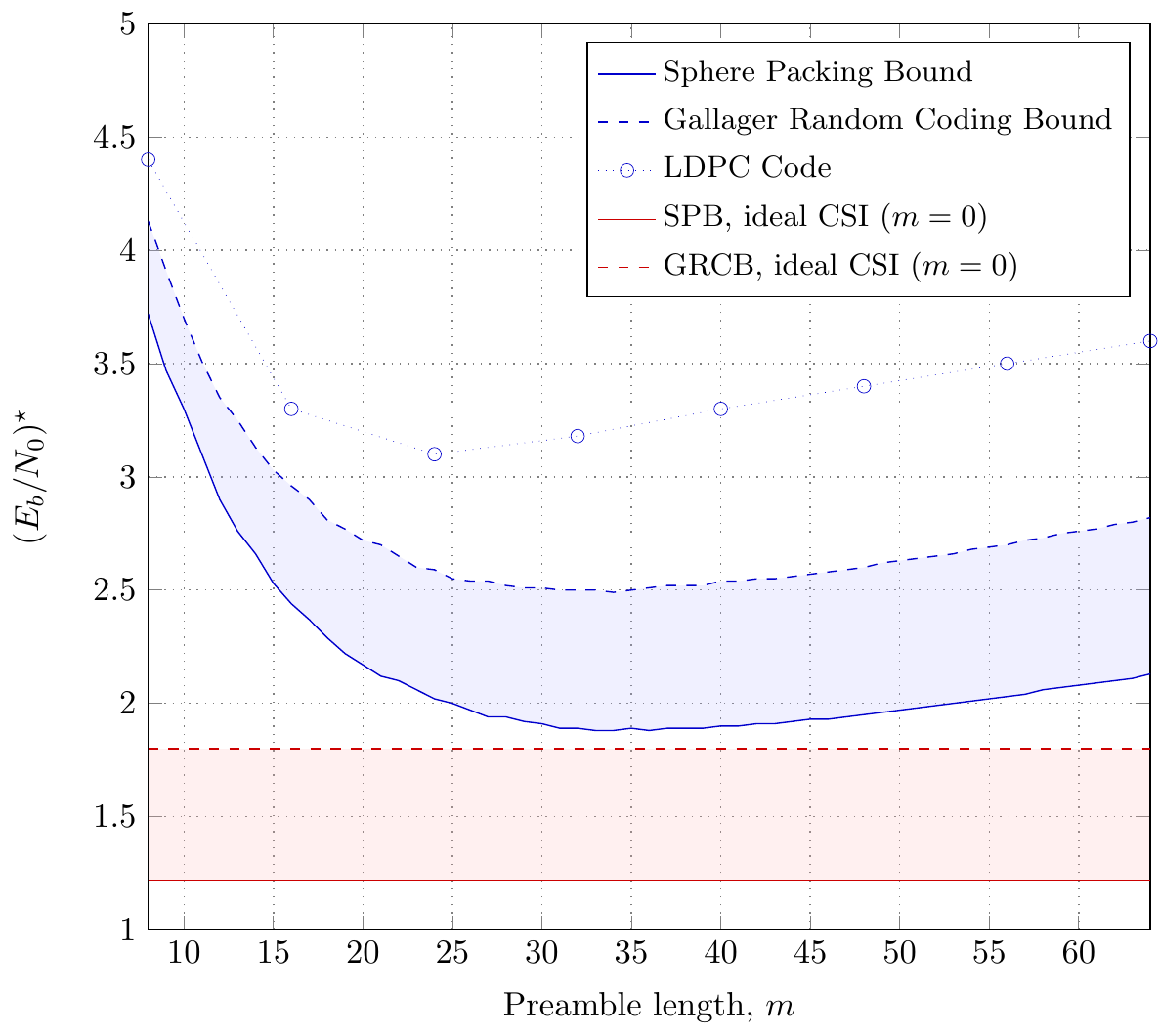}
\end{center}
\caption{Signal-to-noise ratio required to achieve a block error probability of $10^{-3}$ for various preamble lengths. \ac{BPSK}  modulation, frame length $N=512$ symbols, $k=256$ information bits.}	\label{fig:TD_BPSK}
\end{figure}

One may wonder whether changing the length of the preamble in Fig.~\ref{fig:PER_BPSK} results in better performance. 
We address this question in Fig.~\ref{fig:TD_BPSK}, where we depict bounds on the minimum signal-to-noise ratio $(E_b/N_0)^\star$ required to achieve a target error probability of $10^{-3}$ as a function of the preamble length $m$ for fixed $N=512$ and $R=1/2$.
The upper bound is obtained using the \ac{GRCB} whereas the lower bound relies on the \ac{SPB}.
For reference, we also consider the case of ideal CSI, for which $m=0$.
We see from Fig.~\ref{fig:TD_BPSK} that the optimum preamble length is roughly between $24$ and $45$ symbols.
Within this range, the bounds are  flat and give a minimum signal-to-noise ratio of about $2.5$ dB for the \ac{GRCB} (roughly $0.7$ dB away from the ideal \ac{CSI} case), and of $1.9-2$ dB for the \ac{SPB} (again, about $0.7$ dB away from the ideal \ac{CSI} case). The blue shaded area in Fig.~\ref{fig:TD_BPSK} represents the region where the performance of the best $(512-m,256)$ code  lies. 

On the same chart, we provide also the $(E_b/N_0)^\star$ of a specific \ac{LDPC}. 
More specifically, we designed a $(512,256)$ \ac{LDPC} code from an \ac{IRA} ensemble \cite{Jin00:IRA} with degree distribution  pair (edge-oriented) $\lambda(x)=(1/3) x + (2/3) x^3$, $\rho(x)=x^5$.  
We obtained the higher rate codes required to accommodate preambles of different length $m$, by puncturing a corresponding amount of parity bits. 
Specifically, we selected periodic puncturing patterns with period $\lceil(N-k)/m\rceil$. 
A block-circulant version of the \ac{PEG} algorithm \cite{Hu05:PEG} has been used to design the code parity-check matrix. 
The code performance follows closely the prediction of the two bounds, showing an optimum at a preamble length of $24$ symbols. 
At larger preamble lengths, the performance degrades  faster than what predicted by our bounds.
This behavior can be partially explained by the introduction of punctured bits, which may impair the iterative decoding convergence. 
At the optimal preamble length of $24$ symbols, the minimum SNR required by the LDPC code is about $0.5$ dB away from the \ac{GRCB}, which is in good agreement with the results for short binary \ac{LDPC} codes over coherent channels that have been reported in the literature (see e.g. \cite{Liva2016:ShortSurvey,Liva08:QC_GLDPC}). 
Observe that, in this specific case, over-dimensioning the preamble length is less critical than under-dimensioning it.

\begin{figure}
	\begin{center}
	\includegraphics[width=1\columnwidth]{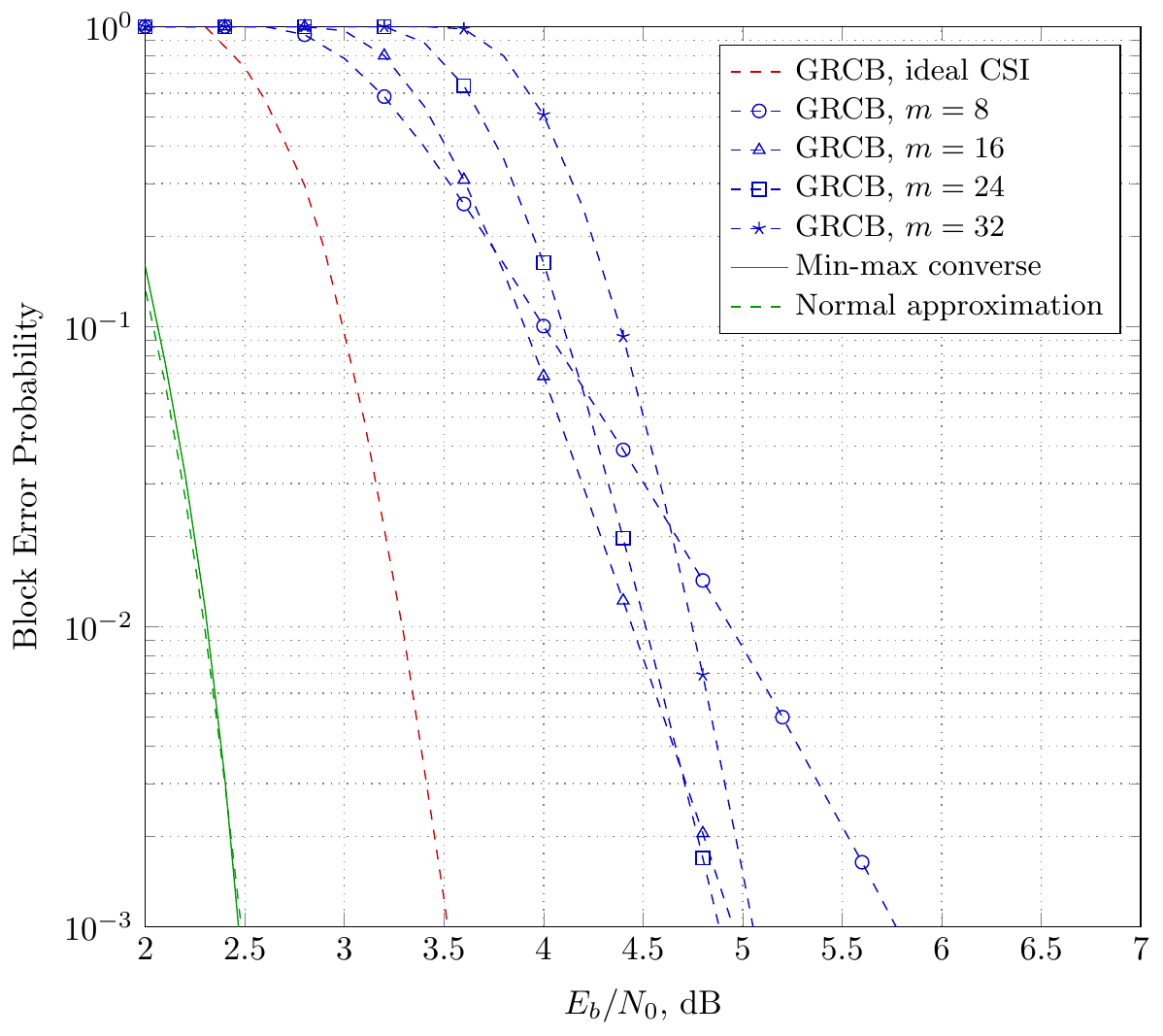}
\end{center}
\caption{Block error probability vs. signal-to-noise ratio according to the \ac{GRCB} with ideal CSI and with pilot-based channel estimation. $16$-QAM, frame length $N=512$ symbols, $k=1024$ information bits.}	\label{fig:PER_QAM}
\end{figure}

\begin{figure}
	\begin{center}
	\includegraphics[width=1\columnwidth]{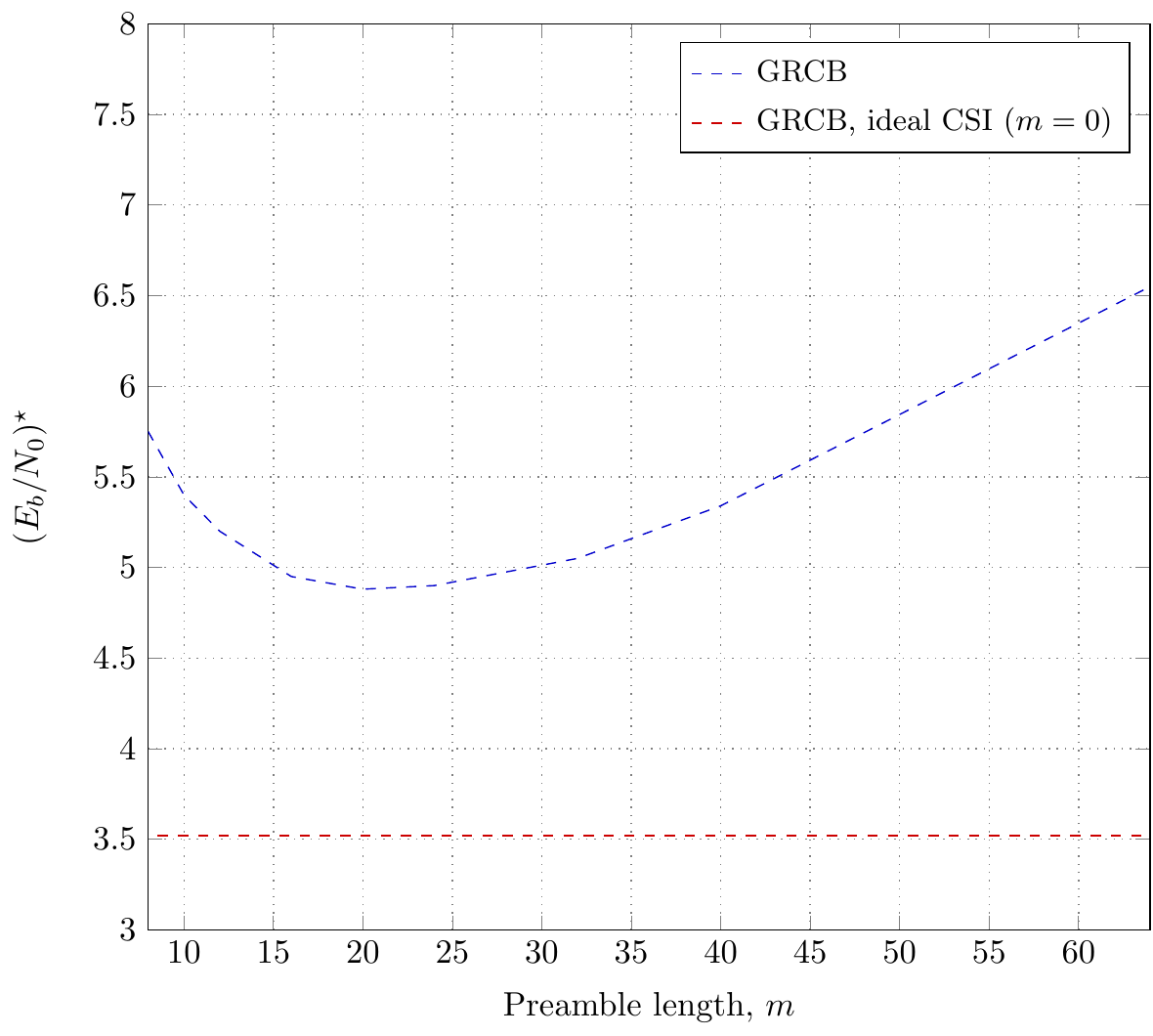}
\end{center}
\caption{Signal-to-noise ratio required to achieve a block error probability of $10^{-3}$ for various preamble lengths. $16$-QAM, frame length $N=512$ symbols, $k=1024$ information bits.}	\label{fig:TD_QAM}
\end{figure}
  
In Figs.~\ref{fig:PER_QAM} and~\ref{fig:TD_QAM}, we provide a similar analysis for the case of a $16$-\ac{QAM} modulation with frame length $N=512$ and $k=1024$ information bits, which result in an overall rate of $R=2$ bits per channel use. 
In Fig.~\ref{fig:PER_QAM}, we plot the \ac{GRCB} for the perfect \ac{CSI} case (here, $N=n$ and $R=R_c$) and for our pragmatic scheme with preamble length $m$ equal to $8, 16, 24,$ and $32$. 
As converse bound, we depict the min-max converse bound~\cite[Th.~27]{Polyanskiy10:BOUNDS} computed for the case of spherical codes, perfect knowledge at the receiver of the amplitude $|h|$ of the channel coefficient. 
Furthermore, we assume that the phase of $h$, which is unknown to the receiver, is uniformly distributed 
All these assumptions yield indeed a converse bound (i.e., a lower bound on the block-error probability), which can be computed following steps similar to the ones reported in~\cite{durisi2016short}.
We also depicted the so-called normal approximation~\cite[Eq.~(296)]{Polyanskiy10:BOUNDS} for the perfect CSI case (i.e., the one for a standard complex AWGN channel).
The small gap between the normal approximation and the min-max converse suggests that the lack of knowledge of the phase of $h$ has a very limited impact on performance if one uses an optimal coding scheme.

We see from Fig.~\ref{fig:PER_QAM} that short preamble lengths yield a better performance at high error probabilities.
However, the gap from the ideal \ac{CSI} case increases rapidly as the error probability decreases. 
The block error probability curves for the case of larger preamble length have a slope that follows the one of the perfect \ac{CSI} case, though with a loss that is due to the use of a larger code rate. 
The gap to the converse bound is more significant than in the BPSK case. 
This is mainly due to the shaping loss (recall that the converse bound relies on spherical codes whereas all achievability bounds assume 16-QAM).

In Fig.~\ref{fig:TD_QAM}, we plot upper bounds (obtained using the \ac{GRCB}) on the minimum signal-to-noise ratio  $(E_b/N_0)^\star$ required to achieve a block error probability of $10^{-3}$, both for the perfect CSI case, and for our pragmatic scheme as a function  of the preamble length. In this case, the optimum preamble length  is about $20$ symbols, which is much shorter than the one in the \ac{BPSK} case. This is due to the larger signal-to-noise ratio at which the $16$-\ac{QAM} scheme operates which allows accurate channel estimation with fewer observations.

\section{Conclusions and Look Forward}
In this paper, we used the mismatched-decoding framework to characterize the fundamental tradeoff occurring in the transmission of  short data packet over an AWGN channel with unknown gain that stays constant over the packet. 
We focused on a pragmatic approach where the transmission frame contains a codeword as well as a preamble that is used to estimate the channel. 
Achievability and converse bounds on the block error probability achievable by this approach are provided and compared with simulation results for schemes employing short low-density parity-check codes. The developed bounds turn out to predict accurately the trade-off between the preamble length and the redundancy introduced by the channel code. 

\bibliographystyle{IEEEtran}
\bibliography{IEEEabrv,RCB_MM}
\end{document}